\documentstyle[preprint,aps]{revtex}

\tightenlines

\begin{document}

\draft
%  \twocolumn[\hsize\textwidth\columnwidth\hsize\csname 
%  @twocolumnfalse\endcsname

\title{\bf A New Finite-lattice study of the Massive Schwinger Model}
\author{P. Sriganesh, C.J. Hamer and R.J. Bursill }
\address{School of Physics,                                              
The University of New South Wales,                                   
Sydney, NSW 2052, Australia.}                      

\date{June 22, 1999}
%\date{\today}

\maketitle 

\begin{abstract}
A new finite lattice calculation of the low lying bound state energies in the massive Schwinger model is presented, using a Hamiltonian lattice formulation.
The results are compared with recent analytic series calculations in the low mass limit, and with a new higher order non-relativistic series which we calculate for the high mass limit.
The results are generally in good agreement with these series predictions, and also with recent calculations by light cone and related techniques.
\end{abstract}                    
\pacs{PACS Indices: 12.20-m, 11.15.Ha}

% \phantom{.}
% ]

\narrowtext
\section{INTRODUCTION}
The Schwinger model\cite{sch,low}, or
quantum electrodynamics in two space-time dimensions, is the simplest
of all gauge theories. It has many properties in
common with QCD, such as confinement\cite{col75}, chiral symmetry
breaking, charge shielding\cite{col75}, and a topological
$\theta$-vacuum\cite{col75,cas,col76}. For these reasons it has become a
standard test bed for numerical techniques designed for the study of QCD,
and has been the subject of intensive study over the years in order to provide some insight into QCD.

	 The purpose in this paper is to carry out a new Hamiltonian
finite-lattice calculation of the `positronium' bound-state energies in
the massive Schwinger model, which can be compared with some recent
analytic perturbation theory calculations. The massless Schwinger model
is exactly solvable, and is equivalent to a theory of free, massive
bosons\cite{sch,low}. A mass perturbation theory
expansion about the zero-mass limit can be performed to treat the
case of low-mass
fermions\cite{car76}. This expansion has recently been carried to second
order by Vary, Fields and Pirner\cite{var} and by Adam\cite{adam}. In the case
of large fermion mass an expansion about the non-relativistic limit can be
performed, which involves a Schr\"{o}dinger
equation with a linear Coulomb potential\cite{ham71}. The next-to-leading
terms in this expansion have been discussed by Coleman\cite{col76}, and are
evaluated in this paper. 

	There have been many numerical calculations of the bound-state
spectrum. In the Hamiltonian lattice formulation, strong-coupling series
were first calculated by Banks, Kogut and Susskind\cite{banks}, extended
by Carroll {\it et al.}\cite{car76}, and have recently been taken to tenth order in
$x = 1/g^{2}a^{2}$ by Hamer, Zheng and Oitmaa\cite{ham97}. 
The series can be reliably extrapolated to the continuum limit, and give
detailed and accurate information about the spectrum, although not quite
as accurate as the finite-lattice method presented in this paper.
Finite-lattice Hamiltonian calculations were performed by Crewther and
Hamer\cite{cre80} and Irving and Thomas\cite{irvtho} over fifteen years ago. Hamer
{\it et al.}\cite{ham97} did some new finite-lattice calculations, but used free
boundary conditions; here we use periodic boundary conditions, which
should give better convergence.

	The model has also been a showcase for the light-front approach
to field theory. An early and quite accurate variational calculation in
the infinite momentum frame was carried out by 
 Bergknoff\cite{ber}. Discrete
Light-Cone Quantization (DLCQ) has been applied to the Schwinger model
by Eller, Pauli and Brodsky\cite{ell},
and produced good results, not only for the lowest states, but for a
wide range of excited states.
Mo and Perry\cite{mo} have used a light-front Tamm-Dancoff
approach using up to four-body states
which produced outstanding results.
A calculation using fermion mass perturbation theory with a variational
 approach was carried out by Harada {\it et al.}\cite{hein} to $O(m^4)$.
 Harada {\it et al.}\cite{har} showed that
the six-body contribution to the lowest-lying states was negligible. 
Kr\"{o}ger and
Scheu\cite{kro98} use a momentum representation with a lattice Hamiltonian
corresponding to a 'fast moving frame' which also gives good results.
Berutto {\it et. al.}\cite{ber} use a improved strong coupling expansion which shows a dramatic improvement of physical parameters compared to previous strong coupling expansions.

In Section II of this paper the continuum formalism and the known analytic 
results for the spectrum 
of the Schwinger model are reviewed. The extension of the
non-relativistic series to order $(g/m)^{5/3}$ is outlined, and the
Hamiltonian lattice formulation is summarized.  
	In Section III the methods of calculation are outlined briefly,
and in Section IV our results are presented. It is our opinion that these are
the most accurate numerical estimates of the low-lying bound-state
energies yet obtained.  Our conclusions are summarized in Section V.

\section{Theory}

\subsection{Continuum Hamiltonian}

The massive Schwinger model is QED in (1+1)D, and can be defined by the
Lagrangian density
\begin{equation}
{\cal L} = - {1\over 4} F_{\mu \nu} F^{\mu \nu} + 
\bar{\psi} (i \not\!\partial - g \not\!\!A - m) \psi
\end{equation}
 where $\psi$ is a two component fermion field,
and
\begin{equation}
F_{\mu \nu} = \partial_{\mu} A_{\nu} - \partial_{\nu} A_{\mu}
\end{equation}
The coupling $g$ in (1+1) dimensions has the dimensions of mass.
Choosing the time-like axial gauge:
\begin{equation}
A_0 =0
\end{equation}
the Hamiltonian density is found to be:
\begin{equation}
{\cal H} =  - i \bar{\psi} \gamma^1 (\partial_1 + i g A_1 ) \psi + m \bar{\psi} \psi
+ {1\over 2} F_{01}^2  \label{hamd}
\end{equation}
The stress-energy tensor $F^{\mu\nu}$ has only one component in one spatial dimension:
\begin{equation}
F^{10}= - \dot{A}^1 = E
\end{equation}

The remaining gauge field component is not an independent degree 
of freedom, but can be eliminated by the equation of motion (Gauss' law):
\begin{equation}
\partial_1 E = - \partial_1 \dot{A}^1 = g \bar{\psi} \gamma^0 \psi
\label{guass}
\end{equation}
Equation (\ref{guass}) determines the electric field E up to a constant
``background" field term
\cite{col76}. The background field will be set to zero for this
study.

In the massless case, the theory has been solved by Schwinger\cite{sch,low},
and becomes equivalent to a theory of free, massive bosons, with
mass
\begin{equation}
{M_1\over g} = {1\over \sqrt{\pi}}\simeq 0.564
\end{equation}

\subsection{Expansions around the massless limit}

For small electron mass $m/g$, one can obtain analytic estimates
by perturbing about the massless limit. Caroll, Kogut,
Sinclair and Susskind\cite{car76} found that the lowest-mass
(``vector'')
state has mass
\begin{equation}
{M_1\over g} = {1\over \sqrt{\pi}} + e^{\gamma} ({m\over g}) + \cdots
\simeq 0.564 + 1.78 ({m \over g}) +\cdots , \label{eq8}
\end{equation}
while the ratio of the next-lowest (``scalar'') mass to the vector mass
was
\begin{equation}
{M_2\over M_1} = 2 - 2 \pi^3 e^{2\gamma} ({m \over g})^2 + \cdots
\simeq 2 - 197 ({m\over g})^2 +\cdots ,
\end{equation}
where $\gamma\simeq 0.5772...$ is Euler's constant.

These results have been extended to second order by Vary, Fields and
Pirner\cite{var} and by Adam\cite{adam}
\begin{equation}
{M_1\over g} = 0.5642 + 1.781 ({m\over g}) + 0.1907 ({m \over g})^2
+ \cdots
\label{10}
\end{equation}
Adam\cite{adam} also found
\begin{equation}
{M_2\over M_1} = 2 - {\pi^3 e^{2\gamma} \over 4} ({m\over g})^2
+\cdots
\simeq 2 - 24.625 ({m\over g})^2 +\cdots
\end{equation}
which differs from the result of Carroll {\it et al.}\cite{car76} by
a factor of $8$.
Hence
\begin{equation}
{M_2\over g} = 1.128 + 3.562 ({m\over g}) - 13.512 ({m \over g})^2
+ \cdots
\label{12}
\end{equation}

\subsection{Non-Relativistic Expansion}
In the non-relativistic limit the bound states can be described by a
Schr\"{o}dinger equation with a linear Coulomb potential\cite{ham71}:
\begin{equation}
\left( {p^2\over m} + {1\over 2} g^2 \vert x \vert \right)
 \Psi (x) = E \Psi (x)\label{nonrel}
\end{equation}
where the non-relativistic energy is,
\begin{equation}
E = M - 2m
\end{equation}
Using reduced variables
\begin{equation}
z = (\frac{mg^{2}}{2})^{1/3}x
\end{equation}
\begin{equation}
\lambda = (\frac{4m}{g^{4}})^{1/3}E ,
\end{equation}
equation (\ref{nonrel}) becomes
\begin{equation}
(\frac{d^{2}}{dz^{2}} - |z| + \lambda)\psi(z) = 0
\end{equation}
which is Airy's equation. The solutions are
\begin{equation}
\psi_n (z) = Ai(z-\lambda_{n})
\end{equation}
where the eigenvalues are given for the symmetric states by
\begin{equation}
Ai'(-\lambda_{n})=0
\end{equation}
and for anti-symmetric states 
\begin{equation}
Ai(-\lambda_{n})=0 .
\end{equation}
Hence the lowest vector (symmetric) state is,
\begin{equation}
{E_1\over g}  \sim  0.642 ({g \over m})^{1/3} \quad {\rm as} \quad m/g \to \infty 
 \label{eq12} \\
\end{equation}
and for the lowest scalar (antisymmetric) state
\begin{equation}
{E_2\over g} \sim 1.473 ({g \over m})^{1/3} \quad {\rm as}\quad m/g \to \infty . 
\end{equation}

Coleman\cite{col76} has calculated the next order terms in the
large-mass expansion. Using his results, the reduced
Schr{\" o}dinger equation becomes
\begin{equation}
\left\{
\left[\frac{d^2}{dz^2} - (|z| - \lambda)\right] - 
\frac{1}{\pi}\left(\frac{2g}{m}\right)^{2/3} + 
\left(\frac{g}{4m}\right)^{4/3}
\left[\frac{d^4}{dz^4} - 2 \delta (z)\right]\right\} \psi (z) =0
\end{equation}
The first extra term is merely a constant self-energy term for the
fermions, while the effect of the remaining terms can be found to
leading order by taking their expectation value with respect to the
leading-order eigenfunctions. By including these terms the results are
\begin{equation}
E_1/g = 0.6418 (\frac{g}{m})^{1/3} - \frac{1}{\pi}\frac{g}{m} -
0.25208
(\frac{g}{m})^{5/3}
\label{24}
\end{equation}
and
\begin{equation}
E_2/g = 1.4729 (\frac{g}{m})^{1/3} - \frac{1}{\pi}\frac{g}{m} +
0.10847(\frac{g}{m})^{5/3}
\label{25}
\end{equation}

\subsection{Lattice Hamiltonian}

Using equation (\ref{hamd}) for the Hamiltonian density, a mapping onto
the lattice can be achieved. The ``staggered" lattice approach of Kogut
and Susskind\cite{kog75} is used. Single component fermion
fields on site $n$ are defined to obey the following anti-commutation rules:
\begin{equation}
\begin{array}{cl}
\{\phi^{\dag}(n),\phi(m)\}&= \delta_{nm}   \\
\{\phi(n),\phi(m)\}&= 0
\end{array}
\end{equation}
The sites $n$ to $n+1$ are connected through a link operator:
\begin{equation}\label{link}
U(n,n+1)=e^{iaeA_1(n)}\equiv e^{i\theta(n)}
\end{equation}
The equivalent lattice Hamiltonian  to equation (\ref{hamd}) is:
\begin{equation}
H = - {i \over 2 a} \sum_{n=1}^N [ \phi^{\dag} (n) e^{i \theta (n)}
\phi (n+1) - {\rm h.c.} ] + m \sum_{n=1}^N (-1)^n \phi^{\dag} (n) \phi
(n)
+ {g^2 a \over 2} \sum_{n=1}^N L^2 (n) \label{latH}
\end{equation}
The number of sites $N$ is even and the correspondence between lattice
and continuum fields is:
\begin{equation}
\phi (n)/\sqrt{a} \to \cases{
\psi_{\rm upper} (x), & $n$ even \cr
\psi_{\rm lower} (x), & $n$ odd \cr }
\end{equation}
and,
\begin{eqnarray}
{1\over ag} \theta (n) &\to& - A^1 (x) \nonumber \\
&& \\
g L(n) &\to& E(x) \nonumber
\end{eqnarray}
The gamma matrices are represented by,
\begin{equation}
\gamma^0 = \left( \begin{array}{rr}
1  &0 \\
0  &-1
\end{array}\right) , \quad
\gamma^1 = \left( \begin{array}{cc}
0  & 1\\
-1 & 0
\end{array}\right)
\end{equation}

A ``compact" formulation has been chosen where the gauge field is an
angular variable on the lattice and $L(n)$ is the conjugate ``spin" variable,
\begin{equation}
[ \theta (n), L (m) ] = i \delta_{nm}
\label{32}
\end{equation}
so that $L(n)$ takes on integer eigenvalues $L(n)=0,\pm 1, \pm 2...$ (which
mimics the quantization of the flux in one dimension\cite{banks} in the
continuum Schwinger model).
Banks {\it et. al.}\cite{banks} have
demonstrated that the lattice Hamiltonian is gauge invariant and reproduces
the Dirac equation and QED in the continuum limit.

The one component fermion fields can be replaced by Pauli spin
operators via a Jordan-Wigner transformation\cite{banks},
\begin{eqnarray}
\phi (n) &=& \prod_{l<n} [ i \sigma_3 (l) ] \sigma^- (n) \\
\phi^{\dag} (n) &=& \prod_{l<n} [- i \sigma_3 (l) ] \sigma^{+} (n)
\end{eqnarray}
giving
\begin{equation}
H =  {1 \over 2 a} \sum_{n} [ \sigma^{+} (n) e^{i \theta (n)}
\sigma^{-} (n+1) + {\rm h.c.} ] + {1 \over 2} m \sum_{n} (-1)^n
\sigma_{3} (n) + {g^2 a \over 2} \sum_{n} L^2 (n) \label{latS}
\end{equation}
Now define the dimensionless operator,
\begin{equation}
W = {2\over ag^2} H = W_0 + x V
\end{equation}
where
\begin{eqnarray}
W_0 &=& \sum_n L^2 (n) + {\mu \over 2} \sum_n (-1)^n \sigma_3 (n) +
N\mu/2 
\label{37} \\
V &=& \sum_n [ \sigma^{+} (n) e^{i \theta (n)} \sigma^- (n+1) + {\rm
h.c.} \nonumber
\end{eqnarray}
For $x \ll 1$ strong coupling perturbation theory can be used on the
model. In the strong coupling limit the unperturbed ground state $\vert
0\rangle$ has,
\begin{equation}
L(n) =0, \quad \sigma_3 (n) = - (-1)^n, \quad {\rm all~} n
\end{equation}
and the operators $exp(\pm i\theta(n))$ act to raise/lower the 
electric field $L(n)$ by one unit, according to equation (\ref{32}).
Using Gauss' law the gauge field can be eliminated,
\begin{equation}
L(n) - L(n-1) = {1\over 2} [\sigma_3 (n) + (-1)^n ]
\end{equation}
Using periodic boundary conditions, $L(N)=L(0)$, there remains one
independent gauge degree of freedom. This is called the ``background"
electric field which results in half-asymptotic particles\cite{col76}.
The ground state energy is the lowest lying eigenvalue, $\omega_0$, in
the sector containing the unperturbed ground state $\vert0\rangle$. The
first excited state, $\omega_1$, is the lowest eigenvalue among the
``vector" states, corresponding in the strong coupling limit to
\begin{equation}
\vert 1 \rangle = {1\over \sqrt{N}} \sum_{n-1}^{N-1}
\left[ \sigma^{+} (n)e^{i\theta(n)} \sigma^- (n+1) - {\rm h.c.} \right] \vert 0
\rangle
\end{equation}
The second excited-state energy $\omega_2$ is
the lowest of a ``band'' of excited states
in the vacuum sector, corresponding in the strong-coupling limit to
the state,
\begin{equation}
\vert 2 \rangle = {1\over \sqrt{N}} \sum_{n=1}^{N-1}
\left[ \sigma^{+} (n) e^{i\theta(n)}\sigma^{-} (n+1) + {\rm h.c.} \right] \vert 0
\rangle ~.
\label{s_sta}
\end{equation}

\section{Method}
The Hamiltonian matrix in the spin representation of equation (\ref{37}) 
has $^{2N}{C}_{N}$ basis states, times the number of background field values allowed. The matrix for each finite N was solved 
by exact
diagonalization using the Conjugate Gradient Method (CGM). 
The basis states were `symmetrized' with respect to translations,
assuming periodic boundary conditions.
 The simulations were run for lattices up to 
$N=22$ sites using approximately 1.5 gigabytes of memory and storing
approximately two million states. The ``vector" and ``scalar" states
can be targeted specifically within the CGM routine. The background field values were taken high enough to achieve convergence to the required accuracy - usually values within $\pm 2$ sufficed.

 Two different
methods were tried to extract the continuum limit.

\subsection*{Method I}
The first method is a conventional double scaling analysis, as
used by Crewther and Hamer\cite{cre80}. Firstly, a sequence of
finite-lattice results is extrapolated to the bulk limit, $N
\rightarrow \infty$, at a fixed coupling $x$ (or lattice spacing $a$).
This was done by means of a sequence extrapolation routine\cite{bar,hen},
namely  the alternating VBS algorithm\cite{vbs}. Errors were estimated by
examining the consistency of the results using different 
VBS parameters $\alpha$. The sequence displays good convergence to the 
bulk limit,
provided the lattice spacing is not too small.
An example is given in Table I.

The estimates of the bulk limit at finite couplings are now
extrapolated to the continuum limit, $a \rightarrow 0$, by performing a
polynomial fit to the data in powers of
$x^{-1/2}$ or $ga$, which conforms to the expected
asymptotic series behavior in the weak coupling limit\cite{cre80}.
Linear, quadratic and cubic fits were made to adjacent sets of data
points, and extrapolated to $a = 0$. The convergence between these
different estimates as the coupling was decreased was used to estimate
confidence limits for the result. An example of the extrapolation is
shown in Figure 3 (see later).
 Note that at small couplings
the estimates of the bulk limit become unreliable and carry large error
bars. 

\subsection*{Method II}

A second method which was tried is a modified version of an approach due 
to Irving
and Thomas\cite{irvtho}. The method involves a simultaneous scaling of
the results with lattice size  $N$ and coupling $ga$, in the fashion of
a ``phenomenological re-normalization group". Eigenvalues are calculated
for a sequence of lattice sizes $N$ and couplings $ga$, constrained by
the requirement
\begin{equation}
Nga=c \qquad \mbox{(a constant)}
\end{equation}
where c is a constant, so that as $N \rightarrow \infty$, $ga \rightarrow 0$, approaching the
continuum in a single limiting process. The limit of the sequence can
again be estimated using the VBS routine. For the massless case,
$m/g = 0$, this procedure works extremely well, as discovered by Irving
and Thomas\cite{irvtho}, and shown in Table II.
For $m/g = 0$, we find by this procedure

\begin{eqnarray}
  M_{1}/g & = 0.56417(2)&  \\
  M_{2}/g &= 1.1284(2)& ,   
\end{eqnarray}
to be compared with the exact results \\
\begin{eqnarray}
  M_{1}/g & = 0.564189583 &  \\
 M_{2}/g & = 1.128379167 & .  
\end{eqnarray}

Irving and 
Thomas\cite{irvtho} in their previous study with $N=14$ were able to match
and even surpass the results obtained with our present Method I. 

For the massive case,however, this procedure turns out to be rather less
successful. 
The reason is not hard to find. For a finite system, we have: 
\begin{equation}
c = Nga=gL 
\end{equation}
where $L$ is the physical length of the system. To get the correct
result for the excited state energies, a box size $L$ will
have to be chosen which is larger than the physical size of the
bound state under consideration. 
For the solvable massless case, the physical size is zero.
But for the massive case, this turns out to
require such large values of $L$ that the extrapolation
to the continuum limit becomes less accurate than Method I. Therefore
this method was  dropped from further consideration.

\section{Results and Analysis}

	Figures 1 and 2 show examples of our estimates of the bulk limit
of the vector and scalar bound-state energies as a function of
lattice coupling $ga$, at small fermion mass ($m/g = 0.125$) and large
fermion mass ($m/g = 16$). It can be seen that the behavior of the
vector energy is monotonic, while the scalar energy shows a
pronounced peak at finite coupling, which moves inwards to smaller
values of $ga$ as ($m/g$) is increased.

	At small $m/g$, the vector energy is almost perfectly linear in 
$y = ga$, and an extrapolation can be made to the continuum limit  
$y = 0$ which is accurate to about 0.5\%. At larger masses there is more
curvature in the data, and any structure moves to smaller values of $y$,
as noted above in Figure 1b).
Figure 3 graphs three different estimates of the continuum limit for 
$m/g = 32$, as functions of y, obtained by making linear, quadratic and 
cubic fits to the data at that value of y. The different estimates 
converge nicely down to $y = 0.1$, where our bulk data points begin to 
lose accuracy, leading to a final estimate of the continuum limit 
accurate to about 1\%.

	For the scalar state, our estimates of the bulk energy do not
 reach far inside 
the peak region before they begin to lose accuracy, as seen in Figures 2a) and 
2b). Thus the extrapolations to the continuum limit are substantially less 
accurate again, at about the 2-3\% level.
 
	Table III lists our estimates of the bound-state energies $E_{1}/g,
E_{2}/g$ in the continuum limit, together with the earlier
finite-lattice estimates of Crewther and Hamer\cite{cre80}, the more recent
light-cone estimates of Mo and Perry\cite{mo}, and the `fast-moving frame'
estimates of Kr\"{o}ger and Scheu\cite{kro98}.
It can be seen that our present estimates are 5-10 times more accurate than 
the old results of Crewther and Hamer\cite{cre80} at small $m/g$, but only 2 times more accurate 
at large $m/g$; while for the scalar state they are 2-5 times more accurate. 
The other authors have not attached error bars to their data. The higher-order 
light cone results of Mo and Perry\cite{mo} are in excellent agreement with 
ours, over the entire range of $m/g$. Those of 
 Kr\"{o}ger and Scheu\cite{kro98} lie a little lower than ours for the vector state at both small and large mass ends of the spectrum.
Harada {\it et al.}\cite{hein} obtain accurate results for low mass using 
a variational low-mass expansion, but because of the $O(m^4)$ errors as the 
results approach $m/g=1$ the errors start to escalate.

	Figures 4 and 5 show graphs of our results as functions of
($m/g$), compared with the analytic perturbation theory predictions
discussed in Section II.
At small $m/g$, the numerical results match on well with the small-mass series 
expansions, equations (\ref{10}) and (\ref{12}), although the scalar series 
begins to diverge away at around $m/g \simeq 0.1$ - presumably further terms in
the series are required to get a reasonable result. At large $m/g$, however, a 
definite discrepancy seems to occur: the numerical results lie 
significantly above the series prediction, by about three standard deviations. 
This discrepancy is somewhat puzzling. A closer look at Figures 1b) and 3),
shows a tendency for the data to curve downwards at small $y$, which may
indicate that our continuum estimates are slightly too high at these large 
values of $m/g$. As for the results of Mo and Perry\cite{mo}, it may be that
the fermion self energy term has not been included in their analysis. 

	The results of Kr\"{o}ger and Scheu\cite{kro98} agree extremely well with the non-relativistic series at large $m/g$. At small $m/g$ they fall below
both the analytic small-mass series and the numerical results of this work and Mo and Perry\cite{mo}. This is presumably because they have not included the
contributions of multi-quark sectors, $qq\bar{q}\bar{q}$ and higher, which 
have significant effect in this region.

	For the scalar state, all results agree well with each other, and with 
the non-relativistic series in the regime of large $m/g$.

\section{Conclusions}
In this work a new finite lattice calculation of the lowest lying bound state 
energies in the massive Schwinger model was carried out. Lattice sizes up to 
$22$ sites were treated using an equal time Hamiltonian lattice formulation. The results were $2-10$ times more accurate than a previous study\cite{cre80}. The non-relativistic series were calculated to a higher order.

These results were compared with the light cone estimates of Mo and Perry\cite{mo} and the ``fast moving frame" estimates of Kr\"{o}ger and Scheu\cite{kro98}. The results generally agree very well, showing light cone techniques are powerful and effective for two-dimensional models, as first demonstrated by Eller {\em et. al.}\cite{ell}. The numerical results are generally in satisfactory agreement with analytic series results at large and small values of $m/g$, except for a small discrepancy in the non-relativistic region for the vector boson. Possible reasons for this discrepancy were discussed.

Calculations of this sort provide a useful demonstration that both lattice gauge theory and light cone techniques can give correct and detailed estimates for the behavior of continuum quantum field theory, in a case where analytic results are available for comparison. The Schwinger model also exhibits some fascinating physical effects in its own right\cite{col75}. In future work, we hope to apply both finite lattice and density matrix re-normalization group (DMRG) techniques to a study of chrial symmetry and background field effects on this model\cite{cre80,kro98}.
\acknowledgments
We would like to thank Dr. Zheng Weihong for his valuable help with this 
project, Dr. Christoph Adam for pointing out an error in an earlier
draft of the paper, and Dr. T. Heinzl for pointing us to reference
\cite{hein}.
We are also grateful to the Australian National University Supercomputer
Facility and the New South Wales Center for Parallel Computing for the
use of their computing resources.
Part of this work was carried out while the authors were visiting the Institute for Theoretical Physics at the University of California, Santa Barbara, and the Center for Nonlinear Studies at Los Alamos National laboratory. We would like to thank both these institutions, and particularly Prof. R. Singh and Dr. J. Gubernatis, for their hospitality.
This research was supported in part by the National Science Foundation under Grant No. PHY94-07194.
It forms part of a research project supported by a grant 
from the Australian Research Council.

%\newpage

\setdec 0.00000000000
\begin{table}
\squeezetable
\caption{An example of convergence to the bulk limit, for
the vector state energy, $E{_1}/g$, at fixed coupling, $ga=0.3$, and
fermion mass, $m/g=1$. The left hand column gives the finite lattice
results as a function of lattice size $N$ running from $4$ to $22$. Subsequent columns give 
sequence extrapolations obtained with the VBS algorithm}
\begin{tabular}{llllll}
N& & & & &\\
4&-0.691618777& & & & \\
6&-0.310147879& 2.369742591 & & & \\
8&0.023788597 & 1.352070697 & 0.558817207& & \\
10&0.290637884 & 0.681116152 & 0.541690825& 0.578989183& \\
12&0.449156794 & 0.517228742 & 0.510023826& 0.509662896&  0.509710838 \\
14&0.496778711 & 0.509548577 & 0.509666964& 0.509718226&  0.509714839 \\
16&0.506848384 & 0.509667696 & 0.509726825& 0.509714344& \\
18&0.509051005 & 0.509706664 & 0.509711057& & \\
20&0.509556263 & 0.509710588 & & & \\
22&0.509674480 & & & & \\
\end{tabular}
\end{table}
\begin{table}
\squeezetable
\caption{Convergence of estimates for the vector state energy, $E{_1}/g$, 
at $m/g=0$ using the Irving-Thomas method (Method II). The left
hand column gives the finite-lattice results as a function of $N$ running from $4$ to $22$.
Subsequent columns give sequence extrapolations obtained with the VBS algorithm}
\begin{tabular}{llllll}
N& & & & &\\
4&0.596129337& & & &  \\
6&0.581212680& 0.571805670& & &  \\
8&0.575443766& 0.569350750& 0.564406265& &  \\
10&0.572480497& 0.568058281& 0.564308303& 0.564213492&  \\
12&0.570706177& 0.567268711& 0.564260122& 0.564177411& 0.564183130 \\
14&0.569535913& 0.566739556& 0.564229677& 0.564184816& 0.564172549 \\
16&0.568710910& 0.566361604& 0.564211540& 0.564216832&  \\
18&0.568100325& 0.566078847& 0.564219012& &  \\
20&0.567631383& 0.565859979& & &  \\
22&0.567260599& & & &  \\
\end{tabular}
\end{table}
\begin{table}
\squeezetable
\caption{Comparison of bound-state energies $E_1/g$, $E_2/g$ as
functions of $m/g$. The finite-lattice estimates obtained in this work
are compared with earlier finite-lattice estimates of Crewther and
Hamer\protect\cite{cre80}, light-cone estimates of Eller
{\it et al.}\protect\cite{ell}
and Mo and Perry\protect\cite{mo}, and the results of Kr\"{o}ger and
Scheu\protect\cite{kro98}.}
\begin{tabular}{rrrrrr}
\multicolumn{1}{c}{} & \multicolumn{1}{c}{Method I}
  & \multicolumn{1}{c}{C \& H}
 & \multicolumn{1}{c}{Eller {\it et al.}} & \multicolumn{1}{c}{Mo \&
Perry} & \multicolumn{1}{c}{Kr\"{o}ger \& Scheu}\\
\multicolumn{1}{c}{$m/g$} & \multicolumn{1}{c}{this work} &
 \multicolumn{1}{c}{\cite{cre80}} & \multicolumn{1}{c}{\cite{ell}}&
\multicolumn{1}{c}{\cite{mo}} & \multicolumn{1}{c}{\cite{kro98}}\\
\hline
\multicolumn{6}{c}{Vector state} \\
0     &  0.563(1)  &    0.56(1)   &         &        & \\
0.125 &  0.543(2)  &    0.54(1)   &  0.58 &    0.54& 0.528 \\
0.25  &  0.519(4)  &    0.52)1)   &  0.53 &    0.52& 0.511 \\
0.5   &  0.485(3)  &    0.50(1)   &  0.49 &    0.49& 0.489 \\
1     &  0.448(4)  &    0.46(1)   &  0.45 &    0.45& 0.445\\
2     &  0.394(5)  &    0.413(5)  &  0.40 &    0.40& 0.394\\
4     &  0.345(5)  &    0.358(5)  &  0.34 &    0.34& 0.339\\
8     &  0.295(3)  &    0.299(5)  &  0.28 &    0.29& 0.285\\
16    &  0.243(2)  &    0.245(5)  &  0.23 &    0.24& 0.235\\
32    &  0.198(2)  &    0.197(5)  &  0.20 &    0.20& 0.191\\
\hline
\multicolumn{6}{c}{Scalar state} \\
0     &  1.11(3)&     1.12(5) &        &      & \\
0.125 &  1.22(2)&     1.11(5) & 1.35   &  1.16& 1.314\\
0.25  &  1.24(3)&     1.12(5) & 1.25   &  1.19& 1.279\\
0.5   &  1.20(3)&     1.15(5) & 1.19   &  1.17& 1.227\\
1     &  1.12(3)&     1.19(5) & 1.13   &  1.12& 1.128\\
2     &  1.00(2)&     1.10(5) & 0.98   &  0.99& 0.991\\
4     &  0.85(2)&     0.93(5) & 0.84   &  0.84& 0.837\\
8     &  0.68(1)&     0.77(5) & 0.69   &  0.70& 0.690\\
16    &  0.56(1)&     0.62(5) & 0.55   &  0.56& 0.559\\
32    &  0.45(1)&     0.49(5) & 0.46   &  0.46& 0.447\\
\end{tabular}
\end{table}
%=======================================================================
\begin{figure}
\caption{Estimated vector state energies $E_1/g$ in the bulk limit ($N \rightarrow \infty$), as a function of coupling $ga$. a) for $m/g=0.125$, 
b) for $m/g=16$. The dashed lines are merely to guide the eye.}
\label{fig:fig1}
\end{figure}
%=======================================================================

%=======================================================================
\begin{figure}
\caption{Estimated scalar state energies $E_2/g$ in the bulk limit ($N \rightarrow \infty$), as a function of coupling $ga$. a) for $m/g=0.125$, 
b) for $m/g=16$.The dashed lines are merely to guide the eye.}
\label{fig:fig2}
\end{figure}
%=======================================================================

%=======================================================================
\begin{figure}
\caption{Convergence to the continuum limit for the vector state energy, 
$E_1/g$, at fermion mass $m/g=32$. Points marked with a circle, square and 
diamond represent linear, quadratic and cubic fits respectively. 
These fits give extrapolated bulk energies at the axis $ga=0$, obtained
from clusters of two,three or four data points around the indicated values
of $y=ga$. The dashed lines give the confidence limits of our final estimate.} 
\label{fig:fig3}
\end{figure}
%=======================================================================

%=======================================================================
\begin{figure}
\caption{Continuum estimates of the vector state energy, $E_1/g$, as a function of $log_2(m/g)$. The open squares represent this work, the diamonds are
results of Mo and Perry\protect\cite{mo}, the triangles represent Kr\"{o}ger
 and Scheu\protect\cite{kro98}. The long
dashed line, dotted line and dot-dash line represent the leading order and
 higher order
non-relativistic series, and an expansion around the massless limit,
 respectively.} 
\label{fig:fig4}
\end{figure}
%=======================================================================

%=======================================================================
\begin{figure}
\caption{Continuum estimates of the scalar state energy, $E_2/g$, as a function
of $log_2(m/g)$.  The open squares represent this work, the diamonds are
results of Mo and Perry\protect\cite{mo}, the triangles represent Kr\"{o}ger
 and Scheu\protect\cite{kro98}. The long
dashed line, dotted line and dot-dash line represent the leading order and
 higher order
non-relativistic series, and an expansion around the massless limit, 
 respectively.}
\label{fig:fig5}
\end{figure}
%=======================================================================

\begin{references}
\bibitem{sch}J. Schwinger, Phys. Rev. {\bf 128}, 2425(1962);
\bibitem{low}J. Lowenstein and J. Swieca, Ann. of Phys. {\bf 68}, 172(1971).
\bibitem{col75}S. Coleman, R. Jackiw and L. Susskind,
   Ann. of Phys. {\bf 93}, 267(1975).
\bibitem{cas}A. Casher, J. Kogut and L. Susskind, Ann. Phys. (N.Y.) {\bf 93}, 267(1975).
\bibitem{col76}S. Coleman, Ann. of Phys. {\bf 101}, 239(1976).
\bibitem{car76}A. Carroll, J. Kogut, D.K. Sinclair and L. Susskind,
Phys. Rev. D {\bf 13}, 2270(1976).
\bibitem{var}J.P. Vary, T.J. Fields and H.J. Pirner, Phys. Rev. D {\bf 53}, 7231(1996).
\bibitem{adam}C. Adam, Phys. Lett. B {\bf 382}, 383(1996); Ann. Phys.
(N.Y.) {\bf 259}, 1 (1997).
\bibitem{ham71}C.J. Hamer, Nucl. Phys. B {\bf 121}, 159(1977).
\bibitem{banks}T. Banks, L. Susskind, and J. Kogut, Phys. Rev. D {\bf 13}, 1043(1976).
\bibitem{ham97}C.J. Hamer, Zheng Weihong and J. Oitmaa, Phys. Rev. {\bf
D56}, 55 (1997).
\bibitem{cre80}D.P. Crewther and C.J. Hamer, Nucl. Phys. B {\bf 170}, 353(1980).
\bibitem{irvtho}A.C. Irving and A. Thomas, Nucl. Phys. B {\bf 215}, 23(1983).
\bibitem{ber}H. Bergknoff, Nucl. Phys. B {\bf 122}, 215(1977).
\bibitem{ell}T. Eller, H.C. Pauli and S.J. Brodsky, Phys. Rev. D{\bf 35}, 1493(1987).
\bibitem{mo}Y. Mo and R.J. Perry, J. Comp. Phys. {\bf 108}, 159(1993).
\bibitem{hein}K. Harada, T. Heinzl and Christian Stern, Phys. Rev. {\bf D57},
2460 (1998)
\bibitem{har} K. Harada, A. Okazaki and M. Taniguchi, Phys. Rev. D {\bf
52}, 2429 (1995).
\bibitem{kro98}H. Kr\"{o}ger and N. Scheu, Phys. Lett. B {\bf
429},58(1998)
\bibitem{ber} F. Berruto, G. Grignani, G. W. Semenoff and P. Sodano, Phys. Rev. {\bf 57}, 5070 (1998)  
\bibitem{kog75}J. Kogut and L. Susskind, Phys. Rev. D {\bf 11},
395(1975)
\bibitem{bar}M.N. Barber and C.J. Hamer, J. Austral. Math. Soc. B{\bf 23}, 229(1982).
\bibitem{hen}M. Henkel and G. Sch\"{u}tz, J. Phys. A {\bf 21}, 2617(1988).
\bibitem{vbs}J.M. Van der Broeck and L.W. Schwartz, SIAM J. Math. Anal. 
{\bf 10}, 658(1979)
%\bibitem{SLAC}S.D. Drell et. al., Phys. Rev. D {\bf 14}, 467,1627(1977)
%\bibitem{Wil74}K.G. Wilson, Phys. Rev. D {\bf 10}, 2445(1974)
\end{references}
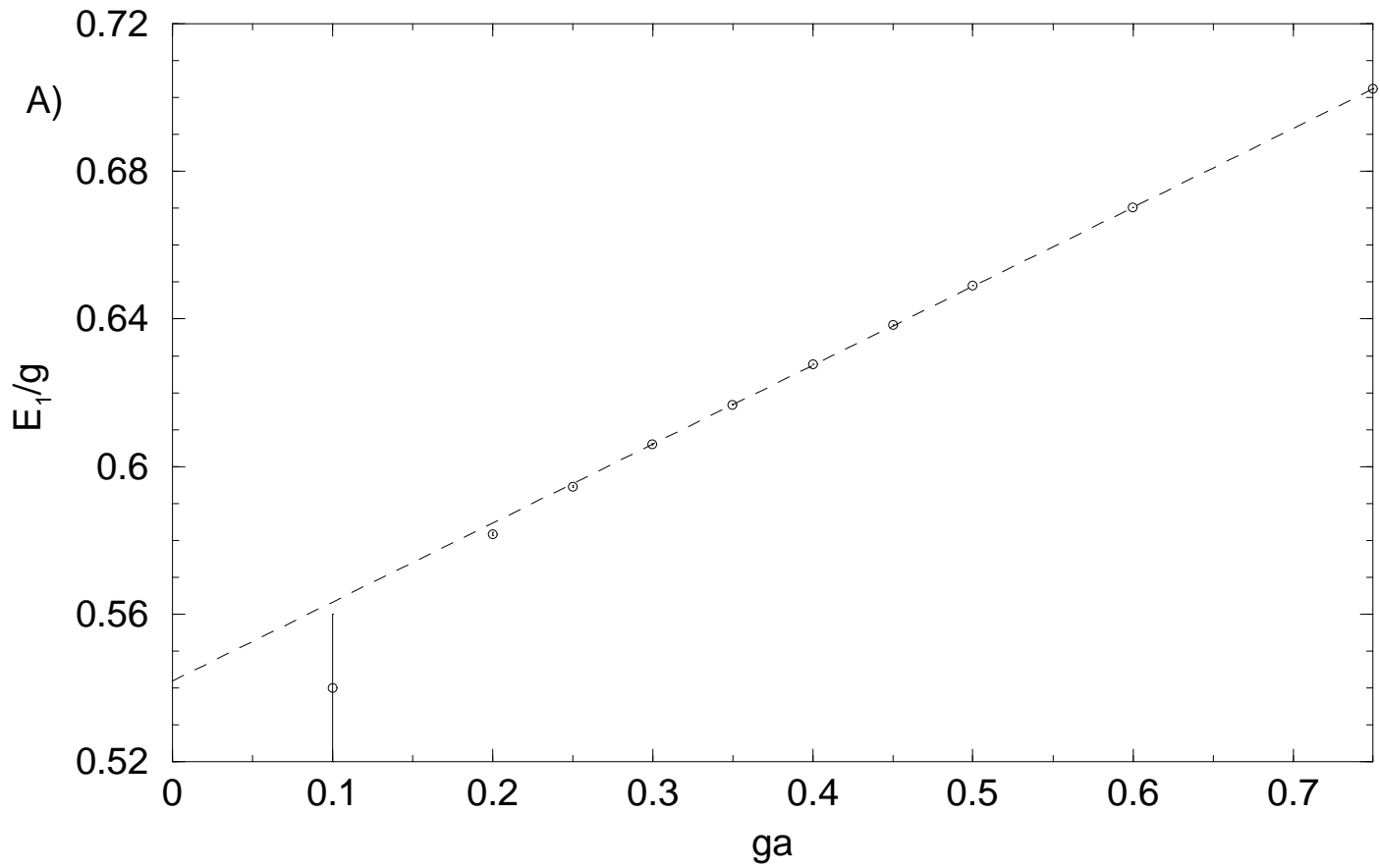
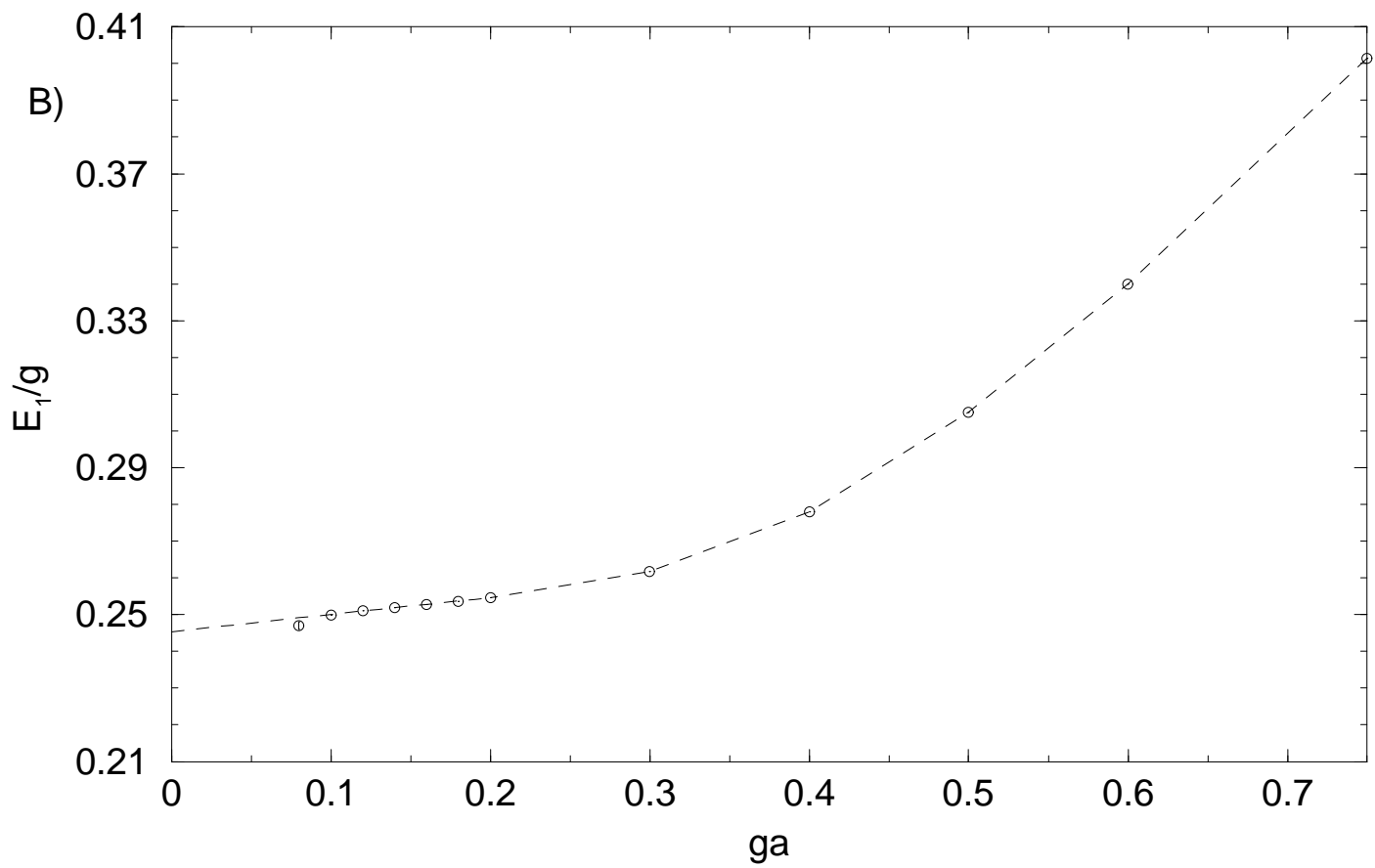
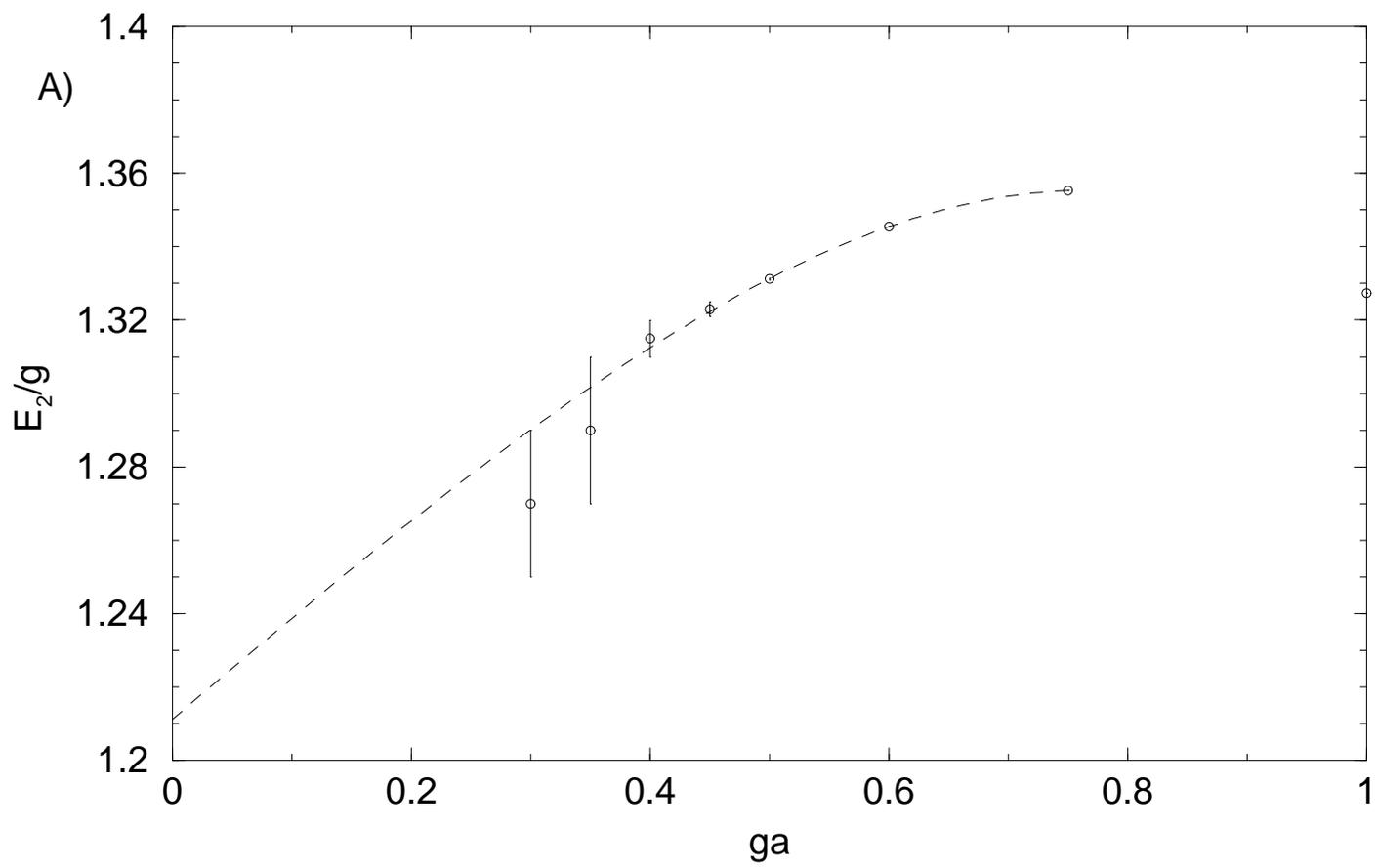
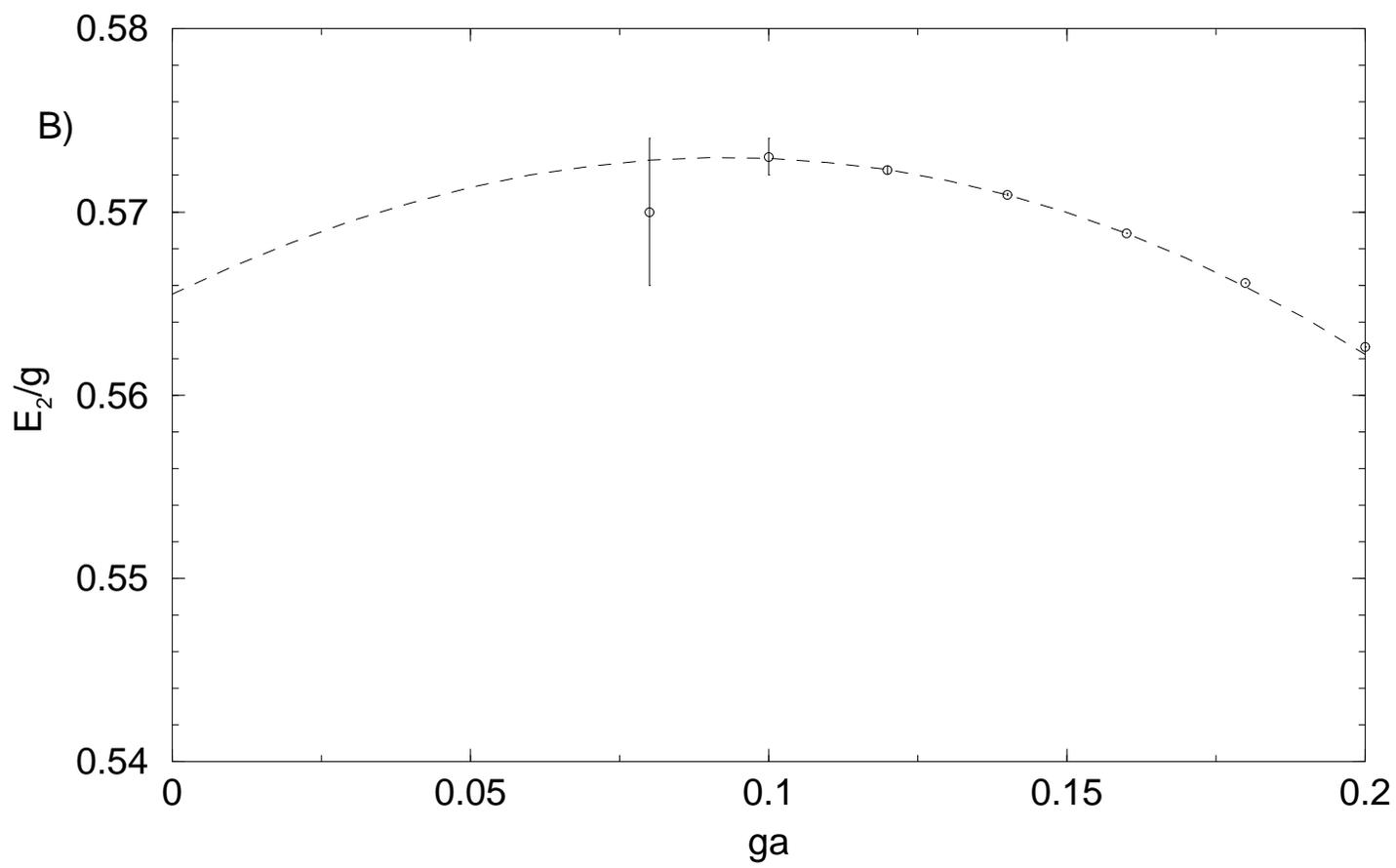
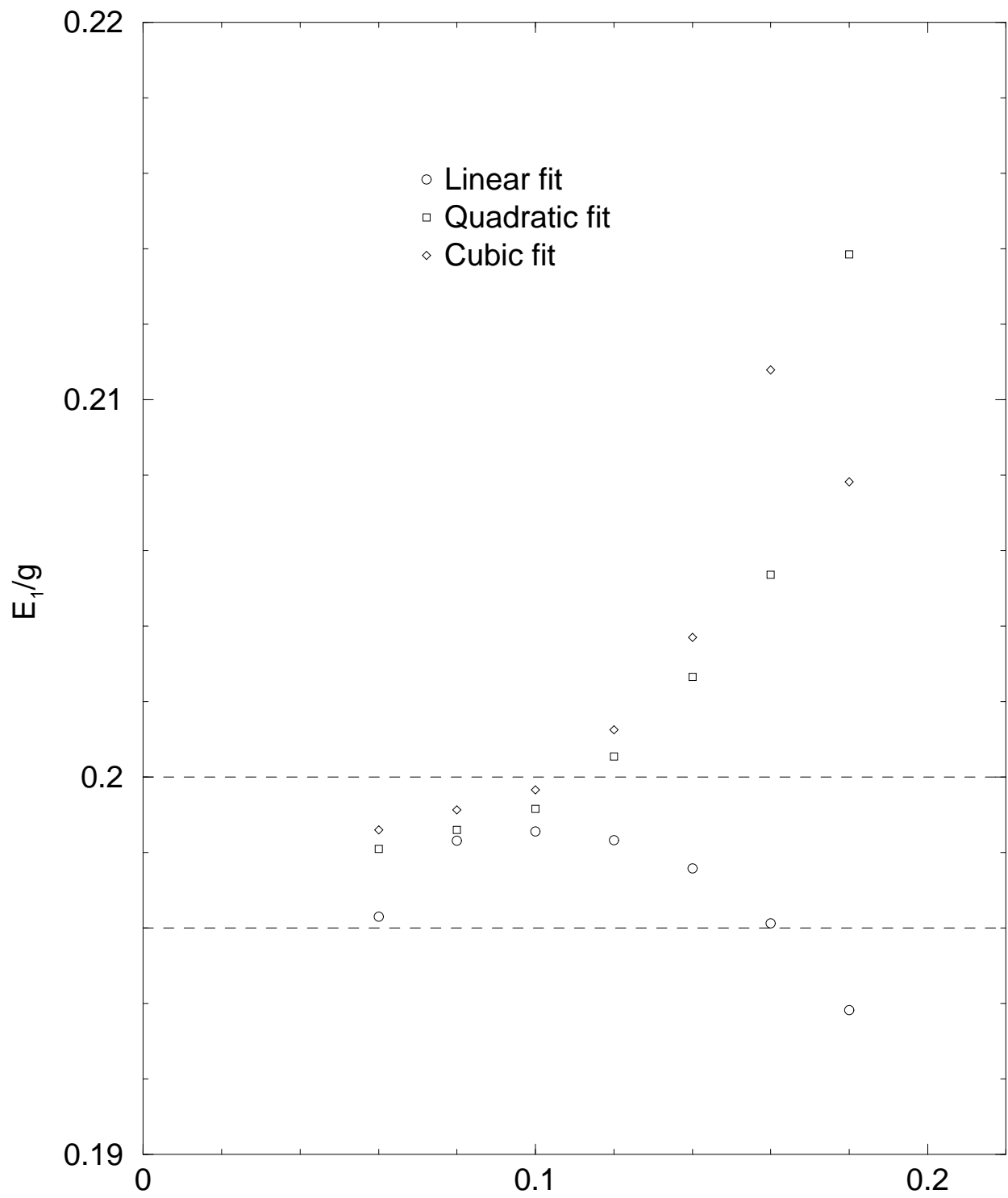
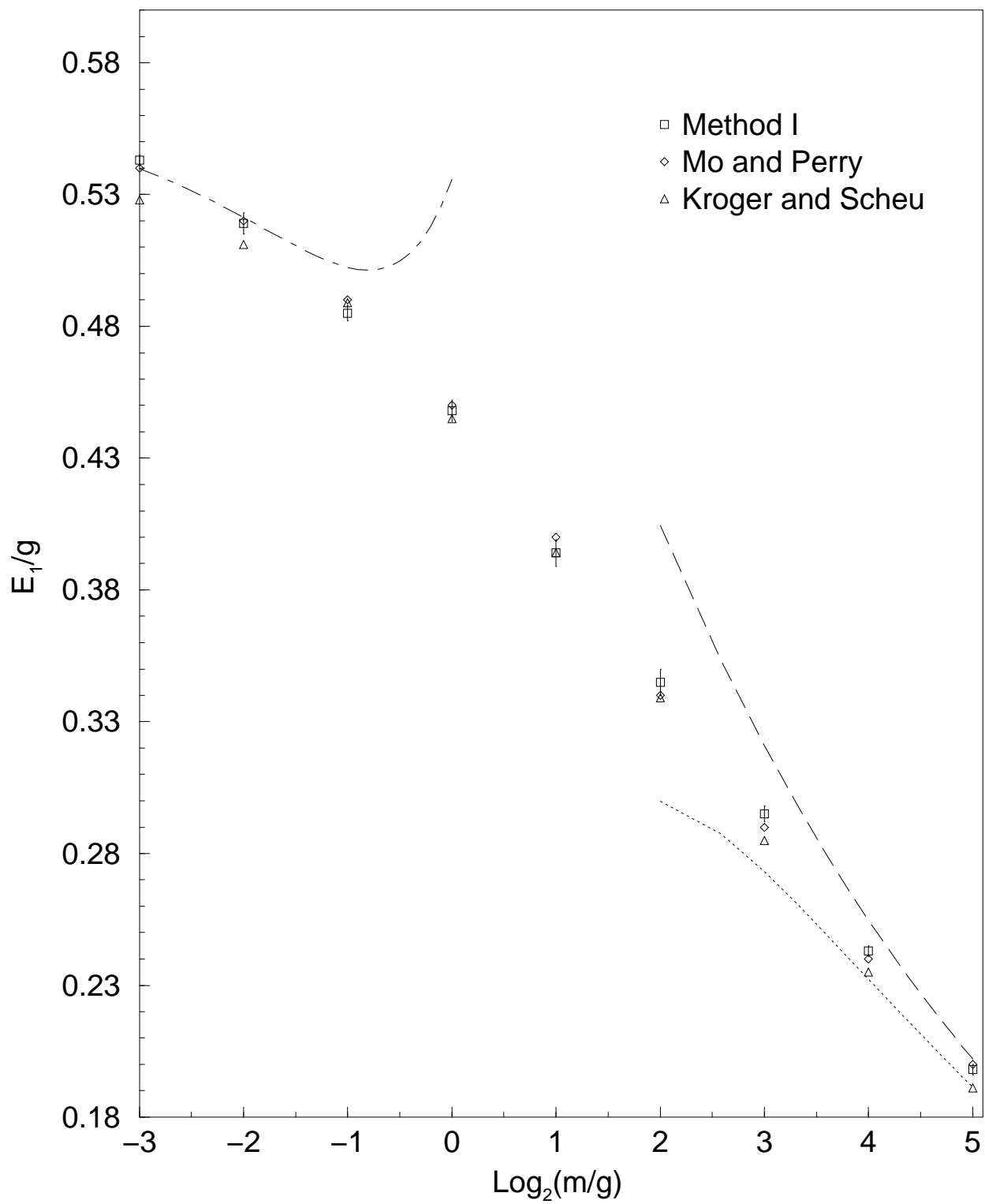
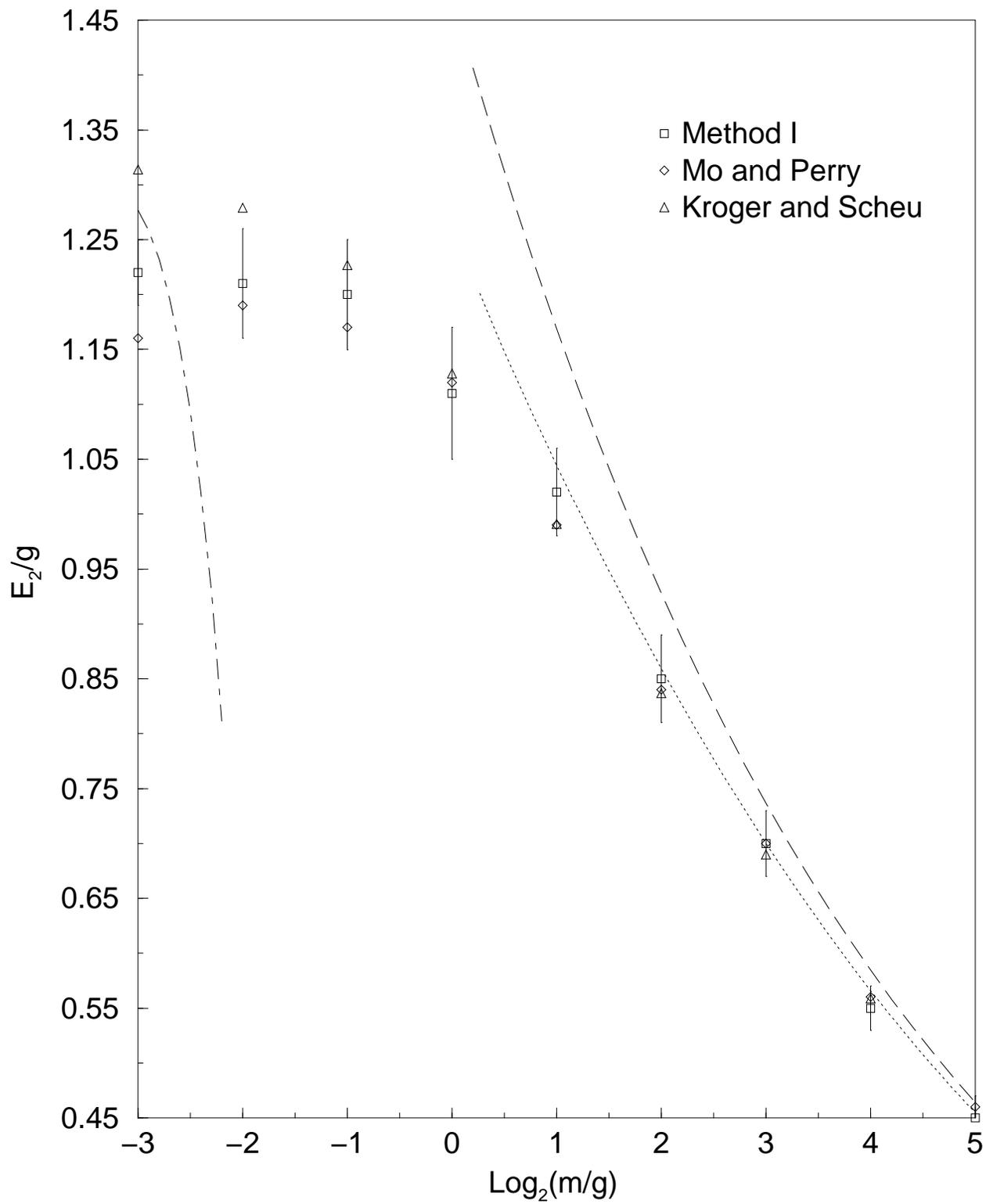
\end{document}